# Enterprise Spreadsheet Management – A Necessary Good

Ralph Baxter
ClusterSeven
10 Fashion Street
London E1 6PX
rbaxter@clusterseven.com, www.clusterseven.com

**ABSTRACT**

This paper presents the arguments and supporting business metrics for Enterprise Spreadsheet Management to be seen as a necessary good. These arguments are divided into a summary of external business drivers that make it necessary and the 'good' that may be delivered to business spreadsheet users involved in repetitive manual processes.

## 1. INTRODUCTION

Four years ago, ClusterSeven was the first company to use the term "Enterprise Spreadsheet Management". We noted that despite the value of spreadsheets to organisations in terms of flexibility, familiarity, speed and cost-efficiency there were notable business deficiencies when they were used as operational applications (i.e. where business practice creates version after version). These deficiencies include the absence of auditability and change control, the difficulty of managing and analysing historical data and the time spent on repetitive manual tasks. In selecting the term "Enterprise Spreadsheet Management" we set out to define a technology sector that addressed these deficiencies, without reducing spreadsheet functionality or performance.

Mark Twain said that 'Work is a necessary evil to be avoided'. As to whether Enterprise Spreadsheet Management should be considered a necessary evil, I will tackle the argument in two steps; firstly to demonstrate that it is 'necessary' and secondly that it is a force for 'good'.

## 2. NECESSARY OR UNNECESSARY?

While the calculation risk inherent in spreadsheets has been extensively discussed since Panko and Halverson (1996), their presence as a systemic part of the operational fabric of businesses has received less research. Croll (2005), Buckner (2004, 2005, 2006) and Pettifor (2005) are notable exceptions. All focus on the pervasive presence of operational spreadsheets and their inherent risks. Moreover, they highlight that businesses using operational spreadsheets typically apply much higher standards of risk and asset management in almost everything else they do – somehow spreadsheets have slipped through the management net. The following changes in business risk and its assessment have now brought this issue to a tipping point in perception.

### 1.1 Volume and value of spreadsheets.

In our experience any material financial services business will have a few hundred to a few thousand operational spreadsheets (excluding any version saves). They range in size from small (1MB or less) to very large (in excess of 150MB). The total volume of operational spreadsheets will commonly exceed 1GB. The most intensive departments (e.g. trading of 'exotic' derivative instruments) have more than 5GB in daily use. Typically each spreadsheet is used over and over again for many months and probably years. This creates multiple version saves with an aggregate data volume many times that of the operational spreadsheets.

This explains why files scans of servers for .xls files in large departments may locate hundreds of thousands and sometimes millions of spreadsheet files.

All of these operational spreadsheets have an impact on the business (as business assets in their own right, as retainers for intellectual property, as consumers of man hours and as a record of business input to decisions). These spreadsheets are often associated with some of the most innovatory and profitable parts of the business (Ashton, 2005). Despite this value, most businesses have no explicit recognition or management of this asset base. Ironically, this is probably because the activity is so prevalent that it has become an unremarkable and implicit part of business.

The spreadsheet asset base within a business clearly requires explicit identification and monitoring as much as any other information resource. For most organisations the scale of the challenge means that manual solutions will be impractical. Enterprise Spreadsheet Management provides a technological solution to the problem, enabling organisations to wrap a monitoring environment around this asset base establishing a basis for all the standard IT policies such as security, change control, disaster recovery and data management.

**1.2 Tactical or Strategic?**

The most common reason for businesses to reject the establishment of a formal enterprise spreadsheet management process is that operational spreadsheets are short-term tactical solutions. As such, there is little reason to invest in managing them, it being better to conserve funds for a future long term solution. This is an example of what Buckner (2003) termed the budget paradox.

As confirmation of this 'denial' it is interesting to note that we have found binary code dating from very early versions of Excel (even Excel 95) in current operational spreadsheets. This longevity suggests clear strategic value.

The reference by Buckner (2006) that some banks are stating that 'spreadsheets are now accepted as strategic solutions' changes this dynamic. The switch in perception from tactical to strategic is a key step in justifying corporate resources (in human, financial and technical terms) to be spent supporting Enterprise Spreadsheet Management.

**1.3 The Expectations of Employees**

Until ten years ago there was an unwritten assumption about computing power: an individual joining a big business would get far more capability inside a business than they would get at home. This has changed – an individual with domestic broadband has enough computing power to do almost anything they want. Now when they join an organisation they will commonly be offered reduced power (e.g. by restrictions on web usage, file downloads, email filtering and through locked-down desk top machines). Much of this restriction comes from concerns regarding security or liability.

The spreadsheet is another example of this end-user power. On the one hand it provides the ability for capable individuals to build enormously powerful models to exploit their understanding of a particular financial environment. At the same time there may be a large business exposure caused by the use of this freedom. Clearly the employer cannot remain ignorant of such exposure and, as with the other end-user applications, decisions must be made as to how it should be managed.

An example of this was the £13.9million fine levied on Citigroup (Conceicao, 2005) where the bank was penalised for pursuing a particular trading strategy. Whilst the use of the spreadsheet was not the reason for the fine, the fact that the bank's technology (both trading systems and spreadsheets) was used to perpetrate these unapproved activities receives specific mention in the list of criticisms.

Enterprise Spreadsheet Management technology delivers an important option to managers faced with this decision – allowing users to retain the accustomed value of spreadsheet speed and flexibility, but monitor what they do.

**1.4 Regulators and Personal Exposure**

One of the greatest changes in the business world in the last decade is the rise in corporate regulation. The Sarbanes-Oxley (SOX) act in the US probably represents the fiercest element in a series of regulatory announcements. If SOX can be boiled down to one sentence it is that it removes the 'I didn't know what was going on' defence for senior executives. When it comes to anything to do with corporate finances that means executives have to know where their numbers came from.

Modern centralised systems pass such transparency tests with ease – audit, security and integrity are part of the design specification. Change control is enforced through IT policies. But for spreadsheets it is a different matter. Who knows what has happened to a spreadsheet? If you are an executive facing incarceration for signing off on these items it becomes crucial that you know who did what. This pressure makes Enterprise Spreadsheet Management an executive necessity and has been the single greatest driver behind the growth in the sector.

All major financial institutions now maintain loss registers to support compliance with the Basel II legislation. Examination of these shows multiple losses (occasionally reaching seven figures) attributable to spreadsheet errors that would have been prevented had Enterprise Spreadsheet Management been in place.

Spreadsheet risk has also become more widely recognised in the general business community. In the last two years there have been regular mentions in the press including the Financial Times (2006), the Wall Street Journal (Gomes, 2006), and the Telegraph (Miller, 2005) as well as numerous trade journals (Laurent, 2006 and Albinus,2006). This has been supplemented by work from Microsoft (2006) who specifically outline a file management approach based on the new version of their product Sharepoint (a file repository). In addition the IT analyst house Gartner have now formally recognised the emergence of a new technology sector called Spreadsheet Control (Heiser, 2006). They anticipate that mainstream adoption will be relatively fast, estimating a period of 2-5 years.

Any future executive defence based on ignorance of the issue is clearly becoming steadily more difficult.

**1.5 What Other Solutions Exist?**
The primary test of necessity is whether there are other solutions that could render a technology-based approach to enterprise spreadsheet management unnecessary. In our experience four alternatives are most commonly suggested:

i. **Forbid the use of business-critical spreadsheets**. Many companies have exactly this policy. This fails for two reasons; the confidence of users in their own capability (and the opportunity to deliver unique insight or make money) means that they do so anyway. Secondly, the transition from non business critical to business-critical usage is not a sharp line; the point at which the policy is transgressed is unclear and hence difficult to enforce.

ii. **Manually document changes** e.g requesting users to self-document changes. The problem with this approach is the level of typical business spreadsheet usage. Furthermore, the desire for speed (that makes spreadsheets so attractive) is counter-productive to manual documentation. Our own experience also shows that users genuinely don't remember all their changes as a 'few' changes always leads to a few more. We have seen examples where a signed commitment to make four specific changes have led to another forty – in the name of 'tidying things up'.

iii. **Prevent changes** by locking the spreadsheet beyond input values. This can be achieved by 'compiling' the spreadsheet into a rigid application or by applying cell and sheet protection. This approach creates two user communities; those without the authority to make changes (typically the beneficiaries of the application) and those with authority (typically the maintainers of the application). When well regimented this process can achieve a first level of control and Enterprise Spreadsheet Management can assist in supporting the integrity of this environment. However, in practice it has two weaknesses: firstly, that the pressure to make changes in time for business deadlines means the authorised community gradually grows to include all the regular business users (thus defeating the attempt to segregate the two communities) and secondly, it fails to confirm that those authorised to make changes have done what was requested (and no more).

iv. **Replace operational spreadsheets** with formal applications. This should be the preferred form of retirement for any mature spreadsheet application with proven business value. However, as noted by PWC (2005) spreadsheets will always be utilised to fill the gap between business needs and installed systems. The materiality of this gap is dependent on many factors such as the rate of evolution of the business, the available IT budget to invest in new systems and the certainty that the business being conducted in spreadsheets merits migration. An example of the latter is that of weather derivatives. A few years ago trading weather derivatives and credit derivatives were both promising new areas of business. Today credit derivatives are everywhere and weather derivatives are rare. Money spent migrating weather derivatives spreadsheets would have been wasted. Replacement strategies therefore have a proper role in the termination of individual spreadsheets, but they do not solve the underlying issue.

**2. GOOD OR EVIL?**

Section 1 focused on the environmental pressures to adopt Enterprise Spreadsheet Management. So it may be necessary, but is it good? Since the perception of whether something is 'good' is largely subjective, I will consider it from the value offered to a typical business user (i.e. not a compliance or technical perspective).

My arguments for user-benefit from Enterprise Spreadsheet Management are derived from seeing how spreadsheets are utilised on a day-to-day basis to support business processes. Despite their speed and flexibility there are many usage aspects within a business environment that are inefficient, leading to user frustration and wasted time. This area of negative user experience represents an attractive target for Enterprise Spreadsheet Management to prove itself a definite good.

**2.1 Version Management**

Many users will have corrupted a spreadsheet and found that the easiest solution is to revert to an earlier copy. In an operational business environment, the problem is much greater. Firstly, users are often working on a spreadsheet that they didn't write – so it is easier to break and more difficult to fix and, secondly, the version to be recovered may not be preserved in hot storage – it often requires a lengthy turnaround for IT to retrieve it from a tape back-up.

With the automatic file and version management contained in an Enterprise Spreadsheet Management solution much of this lost time and frustration can be eliminated. Authorised users can have easy access to any past version through an intuitive timeline interface, without needing to rely on others.

Our own experience in clients suggests that the user time saved simply from provision of a user-friendly automated version recovery utility can justify a material proportion (circa 10%) of the overall costs of Enteprise Spreadsheet Management.

An additional benefit to this approach is that the pattern of operational usage by time and by user becomes apparent, giving an explicit presentation of existing workflow practices surrounding the spreadsheet.

**2.2. Data Integrity**

If the output values from a packaged application change unexpectedly the owner will immediately alert management and investigate the business cause. However, if the output value from an operational spreadsheet changes unexpectedly the first response is to check that the spreadsheet is functioning properly before raising any management alerts. This results in two costs: the overhead of constantly checking of spreadsheets, even when they are correct, plus the delay in addressing any underlying business issue.

A typical business example is the 'product control' or 'middle office' department of investment banks. These departments are essentially the business guardians of trading desk activity. They have the responsibility of validating trader activity and providing aggregated reports of financial positions. When a reconciliation fails they are the ones responsible for uncovering the root cause, be it real or just a spreadsheet error. With spreadsheets containing many tens of sheets and millions of cell values, the process of error discovery, location and repair consumes large amounts of employee time – with the approach essentially a lengthy and frustrating process of "spot the difference".

Investment banks have quoted to us that 30 to 50% of their product controllers' time may be involved in such processes. Enterprise Spreadsheet Management can resolve these questions in seconds rather than hours, removing frustrating aspects of spreadsheet usage and allowing users to focus on added value activities. It can do this by filtering out all of the changes that are part of normal business practice, allowing user to rapidly identify the needles in the haystack.

Enterprise Spreadsheet Management can take the analysis of change events to an additional level, offering richer insight into the business practices that are leading to problems. It allows messaging alerts to be set on cell values to rapidly provide management awareness of a potential problem. For example, an alert might be set to fire if a cell reaches the value of '50', up from a previous value of '40'. On drilling down on this alert the full history of the cell contents will provide additional business context. For example the full data sequence may be **40, 40, 40, 50** or **40, 49, 40, 50** or **20, 30, 40, 50** or **49, 49, 40, 50**. Each of these sequences will suggest different business issues that may need to be addressed.

**2.3 Data Integration**

Operational spreadsheets are not an end in themselves. They are usually part of a longer information supply chain. This re-use of spreadsheet information is achieved either by manual re-keying or by creating spreadsheet links or by direct extraction. All of these methods regularly fail, leading to wasted user hours. Manual re-keying is clearly the least robust – and also invalidates the protection of any previous data validation. Spreadsheet links are notoriously error-prone but are still the preferred way of extracting historical data from spreadsheets e.g. a year-end spreadsheet will link to those of January, February, March etc.

Direct extraction is probably the most robust solution but still subject to failure e.g. where users have entered data outside the expected range.

Enterprise Spreadsheet Management can replace these ad-hoc and non-robust solutions, bringing major user benefits in terms of time saved and uncertainty removed. By creating a centralised, structured and secure database of all validated information this information can be re-used with confidence, using robust, refreshable calls to a secure database. Most importantly, spreadsheet information only needs to be validated once.

**2.4 Analysis of Historical Data**

Successive version saves of an operational spreadsheet provide snapshots of the evolution of the business process that is being supported. Each one is 'final' in the sense that it was a true representation of the state of that part of the business at that time.

The changes in business parameters contained in the spreadsheet therefore offer a highly granular dataset that may be analysed (e.g. for trends, variance and covariance) leading to business insight and potentially new opportunity. Using traditional spreadsheet processes the only ways to access these historical datasets is to open each spreadsheet individually and extract the relevant values (probably into a new spreadsheet), to configure multiple links or to write a specific data extraction component. All of these are highly laborious and prone to error. It certainly does not encourage ad hoc investigation where the inspiration for new insight may be derived.

Enterprise Spreadsheet Management offers a granular time series of the value (and formula) history behind every cell, allowing historical analyses to be performed with ease, thus removing hours of manual activity and leaving user time to focus on the opportunity for added value.

**2.5 Spreadsheet Retirement**

An additional source of user frustration is the timing of spreadsheet retirement. Commonly it is attempted too early, whilst functionality is still immature. As a result the user is called into frequent specification and scope change discussions, damaging the relationship between business and IT.

Enterprise Spreadsheet Management can assist with this process. If the formula history is static, with spreadsheet changes limited only to data values then the spreadsheet is a good candidate for migration. Similarly the spreadsheet can be used as a good specification template. However, if formula change continues to be widespread then any attempt at migration is likely to be beset with scope change issues.

It is perhaps fitting that true Enterprise Spreadsheet Management should assist with eliminating its very requirement – although by then there will be new spreadsheets to manage.

**3. CONCLUSIONS**

Having been responsible for establishing the initial use of the phrase Enterprise Spreadsheet Management we have worked with our clients to establish the quantitative and qualitative returns outlined in this paper. These clients are concentrated in the financial services and commodities trading business environments, where spreadsheet usage is amongst the most intense and complex in the business world. For them Enterprise Spreadsheet Management is not an optional extra, it has become a business necessity. For end-users, the potential 'evil' of an all-seeing big brother can be turned into 'good' by eliminating manual tasks and facilitating other repetitive processes based around spreadsheets.

Blank Page